\documentclass[superscriptaddress,groupedaddress,nofootnoteinbib,12pt]{article} 
\usepackage{graphicx}
\usepackage{color}
\usepackage{dcolumn}   
\usepackage{bm}     
\usepackage{bbm}       
\usepackage{amssymb}  
\usepackage{amsmath}
\usepackage{sectsty}
\usepackage{latexsym}
\usepackage{float}
\usepackage{ifthen}
\usepackage{caption,subfig}
\usepackage{enumerate}
\usepackage{url}
\usepackage{caption,subfig}
\usepackage{jcappub}
\usepackage{amsopn}

\usepackage{amsfonts}
\usepackage{multirow}
\usepackage{array}
\usepackage{booktabs}
\usepackage{rotating}

\usepackage{cancel}

\def\clap#1{\hbox to 0pt{\hss#1\hss}}

\def\({\left(}
\def\){\right)}
\def\[{\left[}
\def\]{\right]}
\def\bea{\begin{eqnarray}}
\def\eea{\end{eqnarray}}
\def\be{\begin{equation}}
\def\ee{\end{equation}}
\def\mpl{M_{\rm Pl}}
\def\d{\mathrm{d}}

\newcommand{\Lag}{\mathcal{L}}

\newcommand{\Tr}{{\rm Tr}}
\newcommand{\diff}{{\rm d}}

\newcommand{\MF}{M}
\newcommand{\MBIs}{M^2}
\newcommand{\Nt}{N_q}

\newcommand{\Pm}{ \hat{P}}

\newcommand{\Id}{\mathbbm 1}

\def\clap#1{\hbox to 0pt{\hss#1\hss}}

\newcommand{\mS}{\mathcal{S}}

\newcommand{\mA}{\mathcal{A}}
\newcommand{\mR}{\mathcal{R}}

\newcommand{\pb}{\bar{p}}
\newcommand{\rhob}{\bar{\rho}}

\newcommand{\Omegah}{\hat{\Omega}}

\definecolor{forestgreen}{rgb}{0.133,0.545,0.133}
\newboolean{editorial}
\setboolean{editorial}{true}
\newcommand{\editorial}[2]{\ifthenelse{\boolean{editorial}}{\textcolor{red}{[\textsf{\textbf{{#1}}}: }\textcolor{blue}{\textsf{{#2}}}\textcolor{red}{]}}{}}

\renewcommand{\d}{\mathrm{d}}
\renewcommand{\vec}[1]{\bm{\mathrm{{#1}}}}

 \def\be   {\begin{equation}}   \def\ee   {\end{equation}}

 \def\ba  {\begin{eqnarray}}   \def\ea  {\end{eqnarray}}

\renewcommand{\thesubsection}{\arabic{section}.\arabic{subsection}}

\begin{document}

\title{Anisotropic deformations in a class of projectively-invariant metric-affine theories of gravity}

\author{Jose Beltr\'an Jim\'enez$^{a,b}$, Daniel de Andr\'es$^b$ and Adri\`a Delhom$^c$}
\affiliation{
$^a$Departamento de F\'isica Fundamental and IUFFyM, Universidad de Salamanca, E-37008 Salamanca, Spain.\\
$^b$Instituto de F\'isica Te\'orica UAM-CSIC, Universidad Aut\'onoma de Madrid, Cantoblanco, Madrid, E-28049, Spain.\\
$^c$Departament de F\'{i}sica Te\`{o}rica and IFIC, Centro Mixto Universitat de
Val\`{e}ncia - CSIC.\\
Universitat de Val\`{e}ncia, Burjassot-46100, Val\`{e}ncia, Spain}

	\emailAdd{jose.beltran@usal.es}
	\emailAdd{daniel.deandresh@estudiante.uam.es}
	\emailAdd{adria.delhom@uv.es}

\abstract{ 
Among the general class of metric-affine theories of gravity, there is a special class conformed by those endowed with a projective symmetry. Perhaps the simplest manner to realise this symmetry is by constructing the action in terms of the symmetric part of the Ricci tensor. In these theories, the connection can be solved algebraically in terms of a metric that relates to the spacetime metric by means of the so-called deformation matrix that is given in terms of the matter fields. In most phenomenological applications, this deformation matrix is assumed to inherit the symmetries of the matter sector so that in the presence of an isotropic energy-momentum tensor, it respects isotropy. In this work we discuss this condition and, in particular, we show how the deformation matrix can be anisotropic even in the presence of isotropic sources due to the non-linear nature of the equations. Remarkably, we find that Eddington-inspired-Born-Infeld theories do not admit anisotropic deformations, but more general theories do. However, we find that the anisotropic branches of solutions are generally prone to a pathological physical behaviour.
}

\maketitle

\section{Introduction}

General Relativity (GR) is still the best candidate to describe the gravitational interactions within the energy range in which gravity has been tested up to date \cite{Will:2014kxa}. From the 1.75 arcseconds in the bending of light during the solar eclipse of 1919 \cite{Dyson:1920cwa,Crispino:2019yew,Crispino:2020txj} to the first observation of the shadow of a super-massive black hole in 2019 \cite{Akiyama:2019cqa} GR has shown an impeccable performance in explaining the measurements related to gravitational physics. Despite this impressive success, there are well grounded reasons to search for departures from GR both at low and high energies (large and microscopic scales). At the lowest energies at which gravity has been tested, the agreement of GR with the observations relies on the existence of dark matter \cite{Persic:1995ru,Bertone:2004pz, Clowe:2006eq,Aghanim:2018eyx} and the inclusion of an unnaturally small cosmological constant in the gravitational action \cite{Aghanim:2018eyx,Bull:2015stt,Riess:1998cb}, which leads to the standard $\Lambda$CDM cosmology. However, even by assuming the existence of dark matter, there could still be tensions concerning observables related to galactic dynamics \cite{McGaugh:2016leg,McGaugh:2000sr,McGaugh:1998tq}, which could very well originate from our poor understanding of structures formation and/or the baryonic physics inside virialised objects. On the other hand, the cosmological evolution also serves as an arena to test the infrarred regime of gravity where new degrees of freedom could dwell (see e.g. \cite{Clifton:2011jh,Joyce:2014kja,Ishak:2018his,Heisenberg:2018vsk}). In the high energy regime, departures from GR are motivated by the search of a UV completion and the avoidance of singularities, which would eventually help to understand the physics at very early times or near the centre of black holes. GR is a perfectly valid effective field theory (EFT) up to (in the most optimistic scenario) the Planck scale \cite{Donoghue:1994dn,Burgess:2003jk,Donoghue:2012zc}, where the EFT approach breaks down and new physics is expected to come in to regularise the theory and render it free of singularities.

Most modifications to GR have been formulated in the metric formalism, where all the (diffeomorphism invariant) effective operators can be written as higher order curvature terms in the action that would correct GR at some high energy scale $\Lambda_G$ (see e.g. the systematic construction of \cite{Ruhdorfer:2019qmk}). Nonetheless, it is known  that at the scale $\Lambda_G$ the theory will enter its strong coupling regime and presents violation of unitarity \cite{Donoghue:1994dn,Burgess:2003jk,Donoghue:2012zc}. One of the very remarkable properties of gravity is its geometrical interpretation and this has motivated  a flare of GR extensions based on extending the geometrical framework where the connection is promoted to an independent entity and treated on equal footing with the metric. A common shortcoming of these so-called metric-affine theories is the general presence of ghost-like degrees of freedom that makes the construction of consistent theories an arduous task \cite{BeltranJimenez:2019acz,Aoki:2019rvi,Percacci:2019hxn,Jimenez:2020dpn}. There is however a particular family of metric-affine theories, commonly dubbed Ricci-Based Gravity (RBG) theories, where the existence of a projective symmetry suffices to render them ghost-free \cite{BeltranJimenez:2019acz,Jimenez:2020dpn}. This is essentially due to this symmetry preventing the appearance of dynamical degrees of freedom associated to the connection and, as a matter of fact, they can be shown to be nothing but GR\footnote{We thank Diego Rubiera for a useful remark on this point.}. These projectively invariant RBG theories comprise for instance the extensively studied cases of $f(\mathcal{R})$ or Eddington-inspired-Born-Infeld theories (see \cite{Olmo:2011uz,BeltranJimenez:2017doy}); and several models within the (projectively invariant) RBG class have already been widely studied both in cosmological \cite{Koivisto:2005yc,Olmo:2005hc,Koivisto:2005yc,Barragan:2009sq,Koivisto:2010jj,Barragan:2010qb,Scargill:2012kg,Olmo:2012vd,Bouhmadi-Lopez:2013lha,Jimenez:2014fla,Bouhmadi-Lopez:2014jfa,Cho:2014xaa,Cho:2014jta,Borowiec:2015qrp,Jimenez:2015jqa,Jimenez:2015caa,Chen:2016nhx,BeltranJimenez:2017uwv,Bouhmadi-Lopez:2017ckh,Pinto:2018rfg,Albarran:2018mpg,Iosifidis:2019jgi} as well as astrophysical scenarios \cite{Olmo:2012nx,Wei:2014dka,Jana:2015cha,Olmo:2015dba,Olmo:2015axa,Olmo:2015bya,Avelino:2015fve,Olmo:2016fuc,Avelino:2016kkj,Olmo:2016tra,Olmo:2017fbc,Menchon:2017qed,Shaikh:2018yku,Shaikh:2018cul,Tadmon:2019tkg,Delhom:2019btt,Olmo:2019qsj,Olmo:2019flu,Afonso:2019fzv,Afonso:2020giy,Rubiera-Garcia:2020gcl}.\\

An appealing feature of RBG theories\footnote{Henceforth we will only consider projectively invariant RBG theories.} is that the independent affine connection turns out to be an auxiliary field that can be integrated out as the Levi-Civita connection of a metric tensor $q_{\mu\nu}$ that can differ from the spacetime metric $g_{\mu\nu}$ as we will review below. As shown in {\it e.g.} \cite{BeltranJimenez:2017doy,Afonso:2017bxr,Afonso:2018bpv,Afonso:2018hyj,Afonso:2018mxn,Delhom:2019zrb}, the definition of the metric $q_{\mu\nu}$ permits an Einstein frame representation with a non-linearly modified matter Lagrangian. In this frame it becomes apparent that the role of the connection is that of an auxiliary field, which effectively encodes new matter interactions with a universal scale that is constrained by experiments \cite{Latorre:2017uve,Delhom:2019wir,LetterBI}. Also, given the Einstein frame representation, solutions can be found by standard methods \cite{Afonso:2019fzv,Afonso:2020giy,Orazi:2020mhb}. The key ingredient to go from the spacetime metric frame to the Einstein frame is the existence of a \textit{deformation matrix} $\Omega^\mu{}_\nu$ relating both metrics as  $q_{\mu\nu}=\Omega_\mu{}^\alpha g_{\alpha\nu}$. This deformation matrix is an on-shell function of the stress-energy tensor and the spacetime metric. As shown below, this matrix is obtained from the connection field equations, which in general are highly non-linear, thus in principle allowing for several solutions. In cosmological applications, it is typically assumed that the deformation matrix has the same symmetries as the energy-momentum tensor and the spacetime metric, so that both metrics share the same symmetries. The existence of this solution is guaranteed by demanding that the non-linear corrections amount to at most a cosmological constant in the low energy limit. 

In this work we will explore the possibility that, due to the non-linearities of the equations that determine the deformation matrix, there could as well exist other solutions for $\Omega^\mu{}_\nu$ which do not respect the stress-energy tensor and the spacetime metric symmetries. In RBG theories, while the matter degrees of freedom evolve in the background given by the spacetime metric, gravitational waves can be associated to perturbations of the metric $q_{\mu\nu}$, and they propagate in the background defined by it \cite{Jimenez:2015caa}. Thus, the possible existence of anisotropic deformation matrices for an isotropic cosmological fluid could introduce interesting effects in gravitational wave propagation that may be worth studying. 

The structure of the paper goes as follows. In section \ref{sec:PITheo} projectively invariant RBG theories are presented, showing the role played by the deformation matrix. In section \ref{sec:AnisoDefSolutions} the conditions for a general theory to admit anisotropic deformation matrices when coupled to an isotropic fluid is studied, particularizing later to EiBI and polynomial theories. The solutions of the general quadratic theory is analysed in greater detail. In section \ref{sec:PhysicalScenarios} we introduce an anisotropic deformation matrix in cosmological and black-hole symmetric scenarios and comment on possible physical implications of the anisotropies. Finally in \ref{sec:AnisoEinstein} we give an account of how the anisotropies are described in the Einstein frame, particularly explaining how to reconcile them with the well-known no-hair theorem in GR cosmologies sourced with a cosmological constant. We then conclude in \ref{sec:Discusion}.

\section{Projectively invariant theories}\label{sec:PITheo}

In this section we will introduce the general formalism for the class of solutions under consideration in this work. As explained above, we will impose a projective symmetry\footnote{This symmetry amounts to the theory being invariant under $\Gamma^\alpha{}_{\mu\beta}\rightarrow \Gamma^\alpha{}_{\mu\beta}+\xi_\mu\delta^\alpha_\beta$ for an arbitrary $\xi_\mu$.} to prevent ghost-like instabilities \cite{BeltranJimenez:2019acz,Jimenez:2020dpn} that we achieve by only considering the symmetric part of the Ricci tensor so the gravitational action can be expressed as
\be\label{RBGaction}
\mS=\frac{1}{2} \mpl^2\MF^2\int\d^4x\sqrt{-g}\Lag(g^{\mu\nu},\frac{1}{M^2}\mR_{(\alpha\beta)}(\Gamma))+\mathcal{S}_m[g_{\mu\nu},\psi]
\ee
with $\Lag$ some scalar function defining the theory, $M$ is a mass scale characterising the deviations from the Einstein-Hilbert (EH) action, and $\mS_m$ is  action for the matter fields $\psi$ that are assumed to be minimally coupled to the spacetime metric for simplicity (the interested reader is referred to \cite{Afonso:2017bxr,Jimenez:2020dpn} for discussions on non-minimally coupled matter in RBG theories). From here on we will omit the dependence on the independent affine connection of $\mR_{\mu\nu}$. Since the Lagrangian must be a scalar built in terms of $g^{\mu\nu}$ and $\mR_{(\mu\nu)}$, it can only depend on the matrix $P^\mu{}_\nu=\frac{1}{\MF^2}g^{\mu\alpha}\mR_{(\alpha\nu)}$. Using now that a general $4\times4$ matrix has four independent scalars that we can choose to be $X_n=\Tr\,\Pm^n$, where a hat denotes matrix notation, the general action can always be expressed as
\be
\mS=\frac{1}{2} \mpl^2\MF^2\int\d^4x\sqrt{-g}F(X_1,X_2,X_3,X_4)+\mathcal{S}_m[g_{\mu\nu},\psi]
\label{action2}
\ee
for some scalar function $F$. Expanding the action at low curvatures leads to
\be
\mS \simeq \frac{1}{2} \mpl^2\int\d^4x\sqrt{-g}\left[\MF^2 F_0+\left.\frac{\partial F}{\partial X_1}\right\vert_0 \mR +\frac{1}{2 M^2}\left(\left.\frac{\partial^2 F}{\partial X_1^2}\right\vert_0 \mR^2+\left.\frac{\partial F}{\partial X_2}\right\vert_0 \mR^{(\mu\nu)}\mR_{(\mu\nu)} \right)+\mathcal{O}\left(\mathcal{R}^3_{(\mu\nu)}\right)\right],
\ee
where the subscript $0$ stands for evaluation at zero curvature. We need to impose $\left.\frac{\partial F}{\partial X_1}\right\vert_0=1$ in order to guarantee that the EH action is recovered at low energies. Notice that the new scale $M$ induces a cosmological constant term for non-vanishing $F_0$. The field equations for the metric and connection derived from the action \eqref{RBGaction} are 
\begin{align}
    &\frac{\partial{F}}{\partial g^{\mu\nu}}-\frac{1}{2}F g_{\mu\nu}=\frac{1}{\mpl^2M^2}T_{\mu\nu},\label{metricfieldeq}\\
    &\nabla_\alpha \left(\sqrt{-g}\frac{\partial{F}}{\partial \mathcal{R}^{\mu\nu}}\right)=0\label{connfieldeq},
\end{align}
with $T_{\mu\nu}$ the stress-energy tensor. The above equation for the connection implies that the independent connection  of this general class of theories is given by the Levi-Civita connection of a metric defined as $\sqrt{-q}q^{\mu\nu}=\sqrt{-g}\partial{F}/\partial \mathcal{R}^{(\mu\nu)}$. Now for Lagrangians which are analytic functions of the Ricci tensor, this metric has an on-shell relation to the spacetime metric $g_{\mu\nu}$ of the form\footnote{This can be easily understood because $F$ is a function of $g^{\mu\alpha} \mathcal{R}_{(\alpha\nu)}$ so that the derivatives with respect to $g^{\mu\nu}$ and with respect to the curvature are related. This further allows to algebraically solve for the derivative in \eqref{connfieldeq} from \eqref{metricfieldeq} in terms of the matter fields and the spacetime metric $g_{\mu\nu}$.}
\be\label{relationqg}
q_{\mu\nu}=g_{\mu\alpha} \Omega^\alpha{}_\nu(g,T),
\ee
where the deformation matrix is an on-shell function of one of the metrics (the relation can be inverted) and the stress-energy tensor. Here $q_{\mu\nu}$ denotes the inverse of $q^{\mu\nu}$. The functional form of the deformation matrix $\Omega^\mu{}_\nu$ depends on the specific theory under consideration, being given by \cite{BeltranJimenez:2017doy}
\be\label{defOmega}
\Omegah\equiv \sqrt{\det \left(\frac{\partial F}{\partial \Pm}\right)}\left(\frac{\partial F}{\partial \Pm}\right).
\ee
Since we are assuming that the Lagrangian is an analytic function of $\Pm^\mu{}_\nu$, we see that the deformation matrix is also an analytic function of $\Pm$. Among other interesting properties, this means that $\Omegah$ commutes with $\Pm$. The on-shell solution for the deformation matrix is obtained from the set of metric field equations, which can be re-written as \cite{BeltranJimenez:2017doy}
\be\label{metriceqsP}
\Pm\frac{\partial F}{\partial\Pm}-\frac12 F\Id=\frac{1}{\mpl^2\MF^2}\hat{T}
\ee
where $\Id$ is the identity matrix and $\hat{T}$ is the matrix representation of  the energy-momentum tensor $T^\mu{}_\nu$. These equations allow to obtain $\Pm$, and consequently the deformation matrix, in terms of the matter variables and the spacetime metric. Notice that while for the EH action the above equations \eqref{defOmega} and \eqref{metriceqsP} give respectively a trivial deformation matrix and the Einstein equations, for other theories they become non-linear. This implies the possibility that neither $\Pm$ nor $\Omegah$ have the same symmetries than the stress-energy tensor and/or the spacetime metric. In particular, though not studied yet in the literature, an isotropic fluid could give rise to anisotropic deformation matrices, the motu of this work being to understand when this possibility can be realized and its implications.

\section{Solutions with anisotropic deformation}\label{sec:AnisoDefSolutions}

After going through the general formalism, let us now focus on the special case when the matter sector is described by a perfect fluid with isotropic pressure. In most of the cases treated in the literature, the isotropy of the energy-momentum tensor is assumed to be inherited by the deformation matrix, which is a reasonable and consistent assumption. Our interest in this work is, however, to go beyond this assumption and explore whether solutions with a deformation matrix  that does not inherit the isotropy of the matter sector are possible. This would imply that the two metrics do not share the same symmetries either. The existence of such solutions is plausible due to the non-linear nature of the equations (were they linear, the symmetries of the energy-momentum tensor must always be inherited by the gravitational sector), in close analogy to the existence of Bianchi I solutions in a universe filled with an isotropic fluid. We will expand on this analogy in Section \ref{sec:AnisoEinstein}. Our Ansatz for the matter energy-momentum tensor and the fundamental matrix $\hat{P}$ will then be
\be
T^\mu{}_\nu=
\left(
\begin{array}{cccc}
  -\rho&   \;& \; &\; \\
  \;&  p & \;  &\;\\
 \; &  \; &p   &\;\\
  \;& \; & \; &p
\end{array}
\right)\quad\text{and}\quad P^\mu{}_\nu=
\left(
\begin{array}{cccc}
  P_0&   &  & \\
  &   P_1&   &\\
  &   &  P_2 &\\
  &  &  &P_3
\end{array}
\right),
\ee
which leads to a deformation matrix of the form $\Omegah=\text{diag}(\Omega_0,\Omega_1,\Omega_2,\Omega_3)$; and also to a simple relation between the $X_n$ and the eigenvalues of $\Pm$, that is $X_n=\sum_{i=0}^3 P_i^n$. The metric field equations \eqref{metriceqsP} read 
\bea
&&P_0\frac{\partial F}{\partial P_0}=\frac12 F-\rhob,\label{eq:P0}\\
&&P_i\frac{\partial F}{\partial P_i}=\frac12 F+\pb,\quad {\rm for}\quad i=1,2, 3;\label{eq:Pi}
\eea
where no summation over $i$ is intended, and we have normalised the density and pressure as $\rhob=\rho/(\mpl^2\MF^2)$ and $\pb=p/(\mpl^2\MF^2)$. We can split the spatial equations (\ref{eq:Pi}) into the isotropic part given by the trace
\be
\frac13\sum_{i=1}^3P_i\frac{\partial F}{\partial P_i}=\frac12 F+\pb
\label{eq:trace}
\ee
and the anisotropic part given by
\be
P_i\frac{\partial F}{\partial P_i}-P_j\frac{\partial F}{\partial P_j}=0,\quad {\rm for}\quad i\neq j
\label{eq:anisotropiccondition1}
\ee
We can alternatively use that the function $F$ can be expressed as $F=F(X_1,X_2,X_3,X_4)$ to re-write the above set of conditions as
\be
\sum_{n=1}^4a_n(P_i^n-P_j^n)=0,\quad{\rm for}\quad i\neq j
\label{eq:anisotropiccondition2}
\ee
with $a_n=n\partial F/\partial X_n$. Out of these three conditions, only two of them are independent because the sum of the three equations identically vanishes. Moreover, since the equations are invariant under permutations of $\Omega_1$, $\Omega_2$ and $\Omega_3$, we can take the two independent conditions as
\bea
a_1(P_1-P_2)+a_2(P^2_1-P^2_2)+a_3(P^3_1-P^3_2)+a_4(P^4_1-P^4_2)&=&0,\nonumber\\
a_1(P_1-P_3)+a_2(P^2_1-P^2_3)+a_3(P^3_1-P^3_3)+a_4(P^4_1-P^4_3)&=&0
\label{eq:anisotropiccondition3}
\eea
From these equations we can easily obtain a set of necessary conditions for the existence of  solutions with a non-isotropic deformation matrix. A remarkable result is that, since these equations do not depend on the matter content, it is only the precise form of the theory what will determine whether anisotropic solutions are possible or not. In particular, it is interesting to look for vacuum anisotropic deformations. The way to proceed then is to solve (\ref{eq:anisotropiccondition3}) for two of the components of $\Omegah$ for the anisotropic branch of solutions (if any) and, then, use (\ref{eq:trace}) and (\ref{eq:P0}) to obtain the full solution with the components of the matrix $\Pm$ in terms of the $\rhob$ and $\pb$.

Obviously, the isotropic solution with $\Omega_1=\Omega_2=\Omega_3$ satisfies (\ref{eq:anisotropiccondition3}). However, given the non-linearity of the conditions, it is possible to have multiple isotropic branches. It is guaranteed by construction that for one of these branches the non-linearities will become irrelevant at low energies. The next non-trivial example is the case with axisymmetry, i.e., two components are equal and different from the third. Without loss of generality we can assume $\Omega_1=\Omega_2\neq\Omega_3$, which implies that $P_1=P_2\neq P_3$.  In that case, the first of the two conditions in \eqref{eq:anisotropiccondition3} is trivially satisfied, but the second one still represents a constraint. In the general case eqs. \eqref{eq:P0}, \eqref{eq:trace} and \eqref{eq:anisotropiccondition3} will also be contraints that should be interpreted as necessary but not sufficient conditions that a particular theory of matter plus gravity has to fulfil in order to admit at least one anisotropic solution. Besides finding non-trivial anisotropic solutions from those equations, one needs to further corroborate that they can be physical, for instance, the resulting $\hat{\Omega}$ must be positive definite. In the following we will illustrate these considerations with some explicit examples.

\subsection{Anisotropic deformations in vacuum}

Let us see whether there is any theory within the projectively invariant RBG class which admits an anisotropic deformation matrix in vacuum. The interest is twofold: 1) because if there is no such theory, all the anisotropic solutions that can be constructed in the presence of matter will not have a well behaved limit at low densities. 2) Because any theory within the RBG class that admits an anisotropic vacuum deformation, since it also admits an isotropic one by construction, will have a nontrivial vacuum structure that could potentially introduce vacuum instabilities. The metric field equations in vacuum are given by \eqref{eq:P0} and \eqref{eq:Pi} with $\bar{\rho}=\bar{p}=0$, which can be written as
\bea
P_\mu\frac{\partial F}{\partial P_\mu}=\frac12 F,\label{eqvacuum}
\eea
where $\mu=0,1,2,3$ and no sumation over $\mu$ is understood here. In general, the above equation implies an on-shell relation of the form $P_0(P_1,P_2,P_3)$. For the particular cases of isotropic ($P_1=P_2=P_3$) and axisymmetric ($P_1\neq P_2=P_3$) this dependence is reduced to $P_0(P_1)$ and $P_0(P_1,P_2)$ respectively. By using the definition of the deformation matrix \eqref{defOmega}, from \eqref{eqvacuum} we also arrive to another on-shell condition that must be satisfied by any vacuum anisotropic solution, that is
\begin{equation}
\frac{\Omega_\mu}{\Omega_\nu}=\frac{P_\mu}{P_\nu} \quad\forall\; \mu,\nu.
\end{equation}
Since we are demanding that all the eigenvalues of $\Omegah$ are positive, the above equation implies that the $P_\mu$'s must all have the same sign when the field equations of the corresponding theory are satisfied. Yet another condition imposed by the positivity of the $\Omega_\mu$'s and the dynamics of RBG is that on-shell
\be
F/P_\mu>0 \quad\forall\;\mu
\ee
must be satisfied, which implies that the Lagrangian must also have the same sign as the $P_\mu$ when the field equations are satisfied. Thus, in principle, an RBG satisfying this conditions could have anisotropic vacuum solutions. Let us now analize particular theories which are of interest by themselves.

\subsection{No anisotropic deformations within EiBI}
One of the most extensively analysed metric-affine theories is the class of EiBI theories (see \cite{BeltranJimenez:2017doy} and references therein). An immediate consequence of the necessary conditions for the existence of solutions with anisotropic deformation matrix expressed in (\ref{eq:anisotropiccondition1}) is that they do not exist for the EiBI theories. In order to see this, we simply need to write (\ref{eq:anisotropiccondition1}) when $F=\sqrt{\det(\Id+\Pm)}$, yielding
\be
\sqrt{\det(\Id+\Pm)}\left(\frac{P_i}{1+P_i}-\frac{P_j}{1+P_j}\right)=0,\quad i\neq j.
\ee
Since $\det(\Id+\Pm)$ must be non-vanishing in order to have a regular deformation matrix, we obtain that the only solution to the above equation is $P_i=P_j$ and, thus, the solution must be isotropic. This result agrees and generalises the findings in the literature. For instance, Bianchi I solutions within the EiBI theory were studied in \cite{Harko:2014nya} and it was found that the deformation matrix was indeed isotropic for an isotropic fluid despite having two Bianchi I Ansatz for $q_{\mu\nu}$ and $g_{\mu\nu}$. The spherically symmetric configurations of EiBI theory coupled to an anisotropic fluid have also been studied in \cite{Menchon:2017qed} with an isotropic deformation matrix. Again, when going to the isotropic case, the obtained solutions for the deformation matrix also become isotropic (in fact, they are proportional to the identity matrix, which is a consequence of having considered a cosmological constant-like fluid).

The result that no anisotropic solutions exist within EiBI gravity can be generalised in a straightforward manner to the functional extensions of the EiBI theory considered in \cite{Odintsov:2014yaa}, where the action is given by an arbitrary function of the scalar $\det(\Id+\Pm)$. In that case, the above condition condition generalises to
\be
F'\det(\Id+\Pm)\left(\frac{P_i}{1+P_i}-\frac{P_j}{1+P_j}\right)=0,\quad i\neq j,
\ee
which again implies the isotropic solution with $P_i=P_j$ provided that the pre-factor in the above equation is non-vanishing. 

\subsection{Theories $f(X_1,X_n)$}\label{sec:FX1Xn}

General results can also be obtained for theories that have a Lagrangian defined in terms of $X_1$ and only one of the higher order scalars $X_n$ with $n=2,3$ or $4$. The presence of $X_1$ is imposed in order to guarantee the existence of one branch of solutions continuously connected with the EH Lagrangian at low curvatures. For these particular cases, the two independent conditions \eqref{eq:anisotropiccondition3} are
\bea
a_1(P_1-P_2)+a_n(P^n_1-P^n_2)&=&0,\nonumber\\
a_1(P_1-P_3)+a_n(P^n_1-P^n_3)&=&0.
\label{eq:anisotropicconditionX1Xn}
\eea
For the axisymmetric case, we can choose $P_2=P_1$ so that the first equation is trivially satisfied, and we have a relation $P_3(P_1)$. For a completely anisotropic solution without axisymmetry, the equations (\ref{eq:anisotropicconditionX1Xn}) imply a relation $P_3(P_1,P_2)$ of the form
\be
\frac{a_n}{a_1}=\frac{P_1^n-P_2^n}{P_1-P_2}=\frac{P_1^n-P_3^n}{P_1-P_3}.
\ee
For $n=2$, this relation can be reduced to $P_1+P_2=P_1+P_3$ which in turn implies $P_2=P_3$ and, consequently, only axisymmetric solutions are allowed. For $n=3$ we instead obtain two branches of solutions, the axisymmetric one, and a second branch with $P_1+P_2+P_3=0$ so the completely anisotropic solutions for $n=3$ must have $\Pm$ with traceless spatial part. Finally, for $n=4$ we again have the axisymmetric branch and possibly another completely anisotropic branch defined by the relation
\be
(P_1^2+P_2^2)(P_1+P_2)=(P_1^2+P_3^2)(P_1+P_3).
\ee
As can be seen by writing the explicit solutions for $P_3$
\be
P_3=-\frac{P_1+P_2\pm\sqrt{-((P_ 1+ P_2)^2+2P_1^2+2P_2^2)}}{2}\quad\text{or}\quad P_2=P_3,
\ee
this equation has no real solutions other than $P_2=P_3$ which is also an axisymmetric solution. Thus, for $n=4$ there can be no completely anisotropic branches. Solving the space of potentially anisotropic solutions for the general case is very cumbersome so in the next section we will focus on the quadratic theory. This will be relevant for the general theories with solutions that are perturbatively close to those of the EH action at low energies so it will be possible to extract information for the general theories from our analysis of the quadratic one.

\subsection{General quadratic theory}\label{sec:GenQuad}

Let us consider the general quadratic theory in terms of the Ricci tensor described by the  function $F(X_{1},X_{2})$ in (\ref{action2})
\be
 F=X_{1}+\alpha X_{1}^{2}+\beta X_{2}=[\hat{P}]+\alpha[\hat{P}]^{2}+\beta[\hat{P}^{2}].
 \label{Fquadratic}
\ee
 Note that the parameter that would have gone with $X_{1}$ is fixed to $1$ in order to recover the EH action at low curvatures. Although this theory may seem to have 2 independent dimensionless parameters $\alpha$ and $\beta$, one of them can be absorbed into the mass scale $\MF^{2}$ and only one parameter remains free (besides the non-linear scale $\MF$). Thus (\ref{Fquadratic}) is the most general quadratic Lagrangian that reduces to the EH in the low curvature limit within the RBG family and captures the perturbative effects of any non-linear theory in that regime. Of course, there could be non-perturbative effects that are not properly captured by \eqref{Fquadratic}, although this would typically imply strong departures from GR in the low energy regime which could be observationally accessible. In order to obtain the dependence of the curvatures $P_{i}$  in terms of the energy content we make use of ($\ref{eq:P0}$) and ($\ref{eq:Pi}$), which particularised for the general quadratic action \eqref{action2} read
\begin{equation}\label{sysquadratic}
    \begin{split}
    &P_{0}-\Tr(\Pm_s)+\alpha\left[3P_0^2+2P_0\Tr(\Pm_s)-\Tr^2(\Pm_s)\right]+\beta\left[3P_0^2-\Tr(\Pm_s^2)\right]+2\rho=0,\\
    &2P_i-\Tr(\Pm)+\alpha\left[4P_i\Tr(\Pm)-\Tr^2(\Pm)\right]+\beta\left[4P_i^2-Tr(\Pm^2)\right]-2p=0,
    \end{split}
\end{equation}
where $i=1,2,3$ and $\Pm_s$ is the spatial $3\times3$ sub-matrix of $\Pm$.  From here on, we will drop the bar in $\bar \rho$ to ease the notation, but all the $\rho$'s appearing in the text should be understood as normalised by $1/(M^2M_{\rm Pl}^2)$. According to what has been discussed in section \ref{sec:FX1Xn},  a quadratic theory can only have isotropic or axisymmetric solutions, but not completely anisotropic solutions are allowed. Thus, in order to look for solutions to the above system of equations (\ref{sysquadratic}) we might first impose isotropy or axisymmetry. In the former case, with $P_1=P_2=P_3$ the above equations \eqref{sysquadratic} reduce to
\begin{align} \label{sysquadraticiso}
\begin{split}
0=&3\alpha\Big[2P_0 P_2-3P_2^2+P_0^2\Big]+3\beta(P_0^2-P_2^2)+(P_0-3P_2)+2\rho\\
0=&P_0+3P_2+3p-\rho;
\end{split}
\end{align}
and in the the axisymmetric case, we find
\begin{align} \label{sysquadraticaxi}
0=&P_0+2P_1+P_2+3p-\rho,\nonumber \\
0=&(P_1-P_2)\Big[1+2\alpha(P_0+2P_1+P_2)+2\beta(P_1+P_2)\Big],\\
0=&\alpha\Big[2P_0 (2P_1+ P_2)-(2P_1+P_2)^2+3P_0^2\Big]+\beta(3P_0^2-2P_1^2-P_2)+(P_0-2P_1-P_2^2)+2\rho,\nonumber
\end{align}
where we have chosen $P_1=P_3\neq P_2$ (note that the physical solutions will not distinguish between this choice and $P_1=P_2\neq P_3$ or $P_1=P_3\neq P_1$). Due to the nonlinearities of the systems,both the isotropic and axisymmetric cases have two branches of solutions. Assuming a barotropic fluid with $p=\omega\rho$, the first isotropic branch (that we will call iso-branch-1) is given by
\begin{align}\label{isobranch1}
\begin{split}
&P_{0}(\rho)=\frac{ (3 \omega -1) (6 \alpha +\beta ) \rho -3 \left(1-\sqrt {1-[4\alpha  (3 \omega -1)+ 2\beta (1+5\omega ) ] \rho +(1-3 \omega )^2 (2 \alpha +\beta )^2\rho ^2  }\right)}{8 \beta }\\
&P_{1}(\rho)=\frac{1+(1-3 \omega) (2 \alpha +3 \beta )\rho-\sqrt {1-[4\alpha  (3 \omega -1)+ 2\beta (1+5\omega ) ] \rho +(1-3 \omega )^2 (2 \alpha +\beta )^2\rho ^2  }}{8 \beta }
\end{split}
\end{align}
and the second isotropic branch (iso-branch-2) is given by the functions
\begin{align}\label{isobranch2}
\begin{split}
&P_{0}(\rho)=\frac{(3 \omega -1) (6 \alpha +\beta ) \rho -3 \left(1+\sqrt {1-[4\alpha  (3 \omega -1)+ 2\beta (1+5\omega ) ] \rho +(1-3 \omega )^2 (2 \alpha +\beta )^2\rho ^2  } \right)}{8 \beta}\\
&P_{1}(\rho)=\frac{1+  (1-3 \omega) (2 \alpha +3 \beta ) \rho +\sqrt {1-[4\alpha  (3 \omega -1)+ 2\beta (1+5\omega ) ] \rho +(1-3 \omega )^2 (2 \alpha +\beta )^2\rho ^2  } }{8 \beta }
\end{split}
\end{align}
For the axisymmetric case, both branches have the same solution for $P_0(\rho)$, namely
\begin{align}
&P_{0}(\rho)=\frac{-2 \rho ^2 (1-3 \omega )^2 (\alpha +\beta ) (2 \alpha +\beta )+2 \rho  (\alpha  (6 \omega -2)+\beta  (5 \omega -1))-1}{4 \beta  \rho  (3 \omega -1) (2 \alpha +\beta )-4 \beta }
\end{align}
and the two branches differ in their solutions for $P_1(\rho)$ and $P_2(\rho)$. The first axisymmtric branch (axi-branch-1) is described by
\begin{align}\label{aniso1}
\begin{split}
&P_{1}(\rho)=\frac{-2 \rho ^2 (1-3 \omega )^2 (\alpha +\beta ) (2 \alpha +\beta )+\rho  (-4 (\alpha +\beta )+12 \alpha  \omega +8 \beta  \omega )-1}{4 \beta  \rho  (3 \omega -1) (2 \alpha +\beta )-4 \beta}\\
&P_2(\rho)=\frac{2 \rho ^2 (1-3 \omega )^2 (2 \alpha +\beta ) (3 \alpha +\beta )+2 \rho  (\alpha  (6-18 \omega )+\beta  (3-7 \omega ))+3}{4 \beta  \rho  (3 \omega -1) (2 \alpha +\beta )-4 \beta}
\end{split}
\end{align}
and the second axisymmetric branch (axi-branch-2) is described by the functions
\begin{align}\label{aniso2}
\begin{split}
&P_{1}(\rho)=\frac{6 \rho ^2 (1-3 \omega )^2 (\alpha +\beta ) (2 \alpha +\beta )^2-4 \rho  (2 \alpha +\beta ) (\alpha  (9 \omega -3)+\beta  (12 \omega -5))+6 \alpha +11 \beta}{12 \beta  \rho  (3 \omega -1) (2 \alpha +\beta )^2-12 \beta  (2 \alpha +\beta )}\\
&P_2(\rho)=\frac{2 \rho  \left(6 \alpha ^2 (3 \omega -1)+\alpha  \beta  (33 \omega -17)+\beta ^2 (15 \omega -7)\right)-6 \rho ^2 (1-3 \omega )^2 (\alpha +\beta )^2 (2 \alpha +\beta )-3 \alpha -5 \beta}{12 \beta  \rho  (3 \omega -1) (\alpha +\beta ) (2 \alpha +\beta )-12 \beta  (\alpha +\beta )}.
\end{split}
\end{align}
As far as the deformation matrix $\hat{\Omega}$ is concerned, it can be written in terms of the $P$'s by means of (\ref{defOmega}). For the general quadratic Lagrangian given by (\ref{Fquadratic}) we find
\begin{equation}\label{Omegaquadratic}
    \Omega_{\mu}=\frac{\left[\prod_{\rho=0}^{3}\left(1+2\beta P_{\rho}+2\alpha \sum_{\gamma=0}^{3}P_{\gamma}\right)\right]^{1/2}}{1+2\beta P_{\mu}+2\alpha(P_{0}+P_{1}+P_{2}+P_{3})}.
\end{equation} 
Let us analyse the behaviour of these solutions for radiation and matter fluids. The first thing to notice here is that while the eigenvalues of $\hat P$, and therefore of $\hat \Omega$, depend on both parameters $\alpha$ and $\beta$ for a matter fluid ($\omega=0$), they do not depend on $\alpha$ for a radiation fluid ($\omega=1/3$), thus $\beta$ is the only relevant parameter that controls the behaviour of radiation fluids, and therefore it can be absorbed into the mass scale $\MF$ (up to a sign) so that the theory is completely determined by this scale in the case of a radiation fluid. Then, while for a radiation fluid $\beta\mapsto-\beta$ is equivalent to $\rho\mapsto-\rho$, for a matter fluid we find an equivalence between $(\alpha,\beta)\mapsto(-\alpha,-\beta)$ and $\rho\mapsto-\rho$. Thus, qualitatively, we have one kind of behaviour for radiation fluids, and two different behaviours for matter fluids, depending on the sign of $\alpha/\beta$.

\subsubsection{Isotropic solutions in the quadratic theory}
\begin{figure}[h]
    \begin{minipage}{\textwidth}
        \centering
        \includegraphics[scale=0.45]{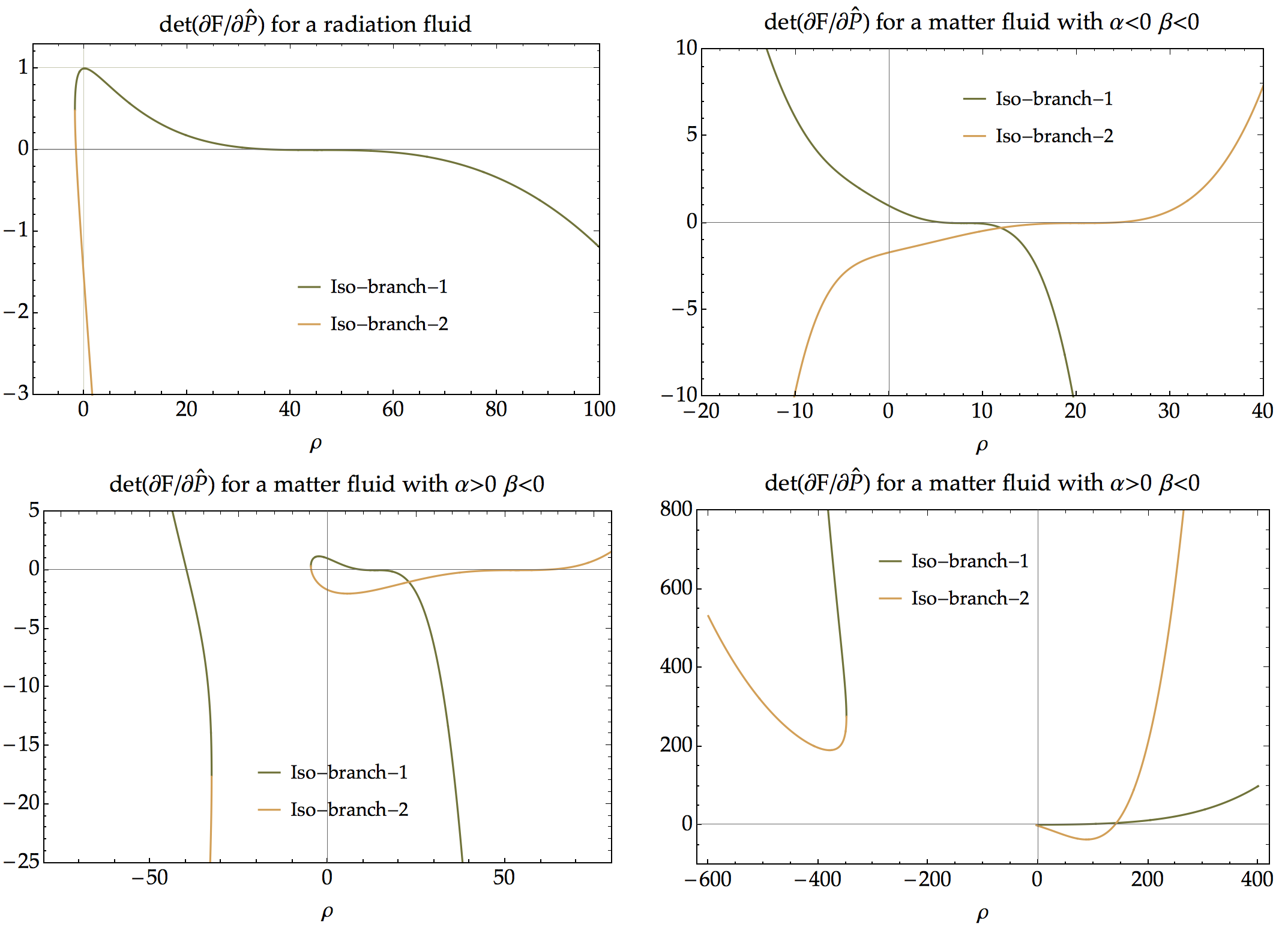} 
    \end{minipage}\hfill
     \caption{The determinant of $\partial F/\partial\hat P$ is plotted for both isotropic branches and  $\beta=-0.1$. The plot above in the right is plotted for $\alpha=-0.01$, and the two below for  $\alpha=0.01$ and $\alpha=0.0345$ (left and right respectively).  It can be seen how $\det(F_{\hat P})=1$ in vacuum for iso-branch-1 in all the cases, but that is never the case for iso-branch-2. } 
    \label{figdetFPiso}
\end{figure}

Isotropic solutions (figure \ref{figdetFPiso}) have already been studied in \cite{Barragan:2010qb} where asymptotically Minkowski solutions and bouncing solutions are found. We will here review the behaviour of the deformation matrix for these solutions. Given that $\Omegah$ is proportional to the square root of $\det(F_{\hat P})$, we must first study the sign of $\det(F_{\hat P})$ for the different solutions (we will assume $\beta<0$). The qualitatively distinct cases for $\det(F_{\hat P})$ in isotropic solutions are plotted in fig.\ref{figdetFPiso}. 

For a radiation fluid we have that $\det(F_{\hat P})$ is positive in the interval $\rho\in(\frac{3}{16\beta},-\frac{9}{2\beta})$ and negative for $\rho>-\frac{9}{2\beta}$ in the iso-branch-1 (given by \eqref{isobranch1}); and it is positive in the interval $\rho=(\frac{3}{16\beta},\frac{1}{6\beta})$ and negative for $\rho>\frac{1}{6\beta}$ in the iso-branch-2 (given by \eqref{isobranch2}). At $\rho=\frac{3}{16\beta}$ both branches give the same value for $\det(F_{\hat P})$, and it becomes complex (in both branches) for $\rho<\frac{3}{16\beta}$. Thus the two branches come from one single solution in the complex plane. For the matter dominated case the analysis is a bit more complex in the general case. We find two different qualitative behaviors that depend on the relative sign between $\alpha$ and $\beta$. For the case with $\alpha$ and $\beta$ having the same sign, we see that each both branches have a zero in $\det(F_{\hat P})$ at positive values of $\rho$. Indeed the zeros are given by $\rho=\frac{-\sqrt{3 \alpha  \beta +\beta ^2}-6 \alpha -2 \beta }{12 \alpha ^2+7 \alpha  \beta +\beta ^2}$ in the iso-branch-1 and $\rho=\frac{\sqrt{3 \alpha  \beta +\beta ^2}-6 \alpha -2 \beta }{12 \alpha ^2+7 \alpha  \beta +\beta ^2}$ in the iso-branch 2. In this cases both branches have $\det(F_{\hat P})\in\mathbb{R}$ for all values of $\rho$, with the iso-branch-1 monotonically decreasing, and the iso-branch-2 monotonically increasing. Only the iso-branch-1 satisfies that $\det(F_{\hat P})=1$ in vacuum, thus recovering GR. The case with opposite signs of $\alpha$ and $\beta$ is much more involved. This is due to the fact that there are more possible values of $\rho$ at which $\det(F_{\hat P})$ has zeroes or poles, as well as intervals in which it becomes complex; and this in general depends on the particular values of the parameters $\alpha$ and $\beta$. In fig.\ref{figdetFPiso} we plotted two of the possible cases. Note that in these cases, the richer structures of zeros and plots of $\det(F_{\hat P})$ gives rise to disconnected (in a continuity sense) sub-branches within the two isotropic branches. Each of the sub-branches of one of the branches always connects smoothly with one of the sub-branches of the other branch, thus implying again that both branches come from a unique solution in the complex plane.

\begin{figure}[h]
		\centering
		\begin{minipage}{\textwidth}
        \centering
        \includegraphics[scale=0.45]{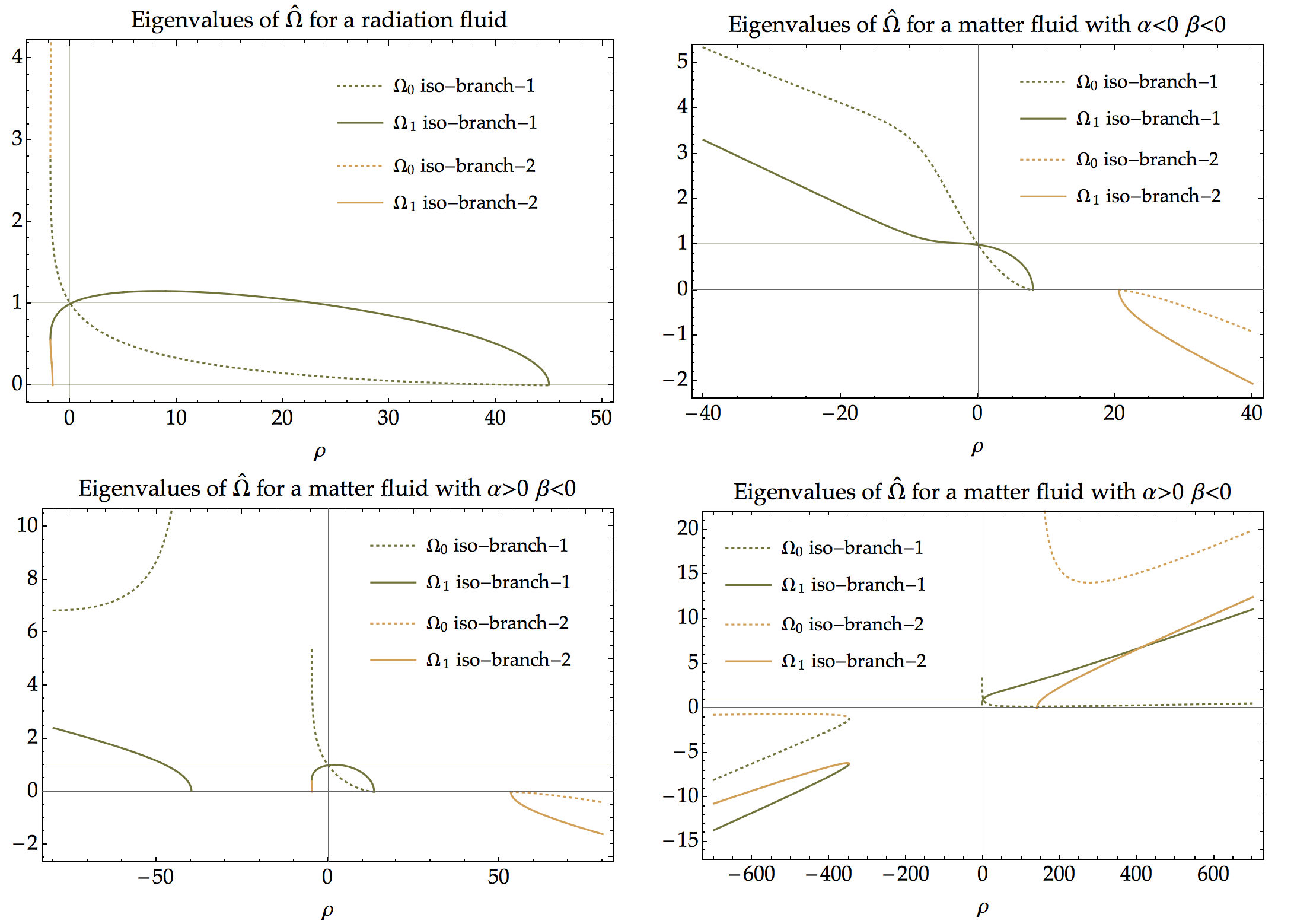} 
    \end{minipage}\hfill
       \caption{The eigenvalues of the deformation matrix are plotted for both isotropic branches and  $\beta=-0.1$. The plot above in the right is plotted for $\alpha=-0.01$, and the two below for  $\alpha=0.01$ and $\alpha=0.0345$ (left and right respectively).  It can be seen how the deformation matrix reduces to the identity in vacuum for iso-branch-1 in all the cases, but that is never the case for iso-branch-2. } 
    \label{fig:Omegaiso}
\end{figure}

A feature worth to note is that, for isotropic branches, the value of $\det(F_{\hat P})$ in vacuum is independent of $\alpha$ and $\beta$, and it evaluates to $1$ for iso-branch-1 and to $-27/16$ for iso-branch-2. Given that the deformation matrix is proportional to $\sqrt{\det(F_{\hat P})}$, the iso-branch-2 does not have a well defined Einstein frame in vacuum. Regarding the properties of the deformation matrix , the first thing to point out is that the value of the deformation matrix in vacuum for isotropic solutions does not depend on the values of the parameters $\alpha$ and $\beta$, and it is the identity for iso-branch-1, whereas for iso-branch two we find $\Omegah_{\rho\to0}=i\sqrt{3}/2\; {\rm diag}(-3,1,1,1)$. This implies that while for iso-branch-1 the non-linearities always smoothly disappear vacuum, the Einstein frame of iso-branch-2 is not well defined in vacuum since there are no real solutions in this case. This properties can be verified in fig.\ref{fig:Omegaiso}, where we plot the eigenvalues of the deformation matrix for the different cases. From the plots we can also see how, except for the radiation solutions, matter solutions with $\alpha/\beta>0$ and one of the subcases of matter solutions with $\alpha/\beta<0$ (corresponding to $3\alpha+\beta<0$ ), the deformation matrix becomes singular at some maximum density thus jeopardising the construction of the Einstein frame at higher densities. Physically, this is associated to an actual upper bound for the energy density allowed in these branches of the theory, a property with the potential to regularise both black hole and cosmological solutions and, consequently, the avoidance of singularities by generating a wormhole throat or a bounce when the energy densities reach this critical value \cite{Koivisto:2005yc,Barragan:2009sq,Koivisto:2010jj,Barragan:2010qb,Scargill:2012kg,Olmo:2012nx,Olmo:2015dba,Olmo:2015bya,Olmo:2016fuc,Olmo:2016tra,Olmo:2017fbc,Menchon:2017qed}. It is important to stress however that these solutions can also present other pathologies (instabilities, violations of energy conditions, superluminalities, etc.). Notice that this does not happen for the $3\alpha+\beta > 0$ subcase of the $\alpha/\beta>0$ solutions\footnote{The sign of $3\alpha+\beta$ is related to the structures of zeroes of $\det(F_{\hat P})$. }, where the deformation matrix does not become critical at any positive value of $\rho$.

\subsubsection{Axisymmetric solutions in the quadratic theory}

\begin{figure}[h]
		\begin{minipage}{\textwidth}
        \centering
        \includegraphics[scale=0.42]{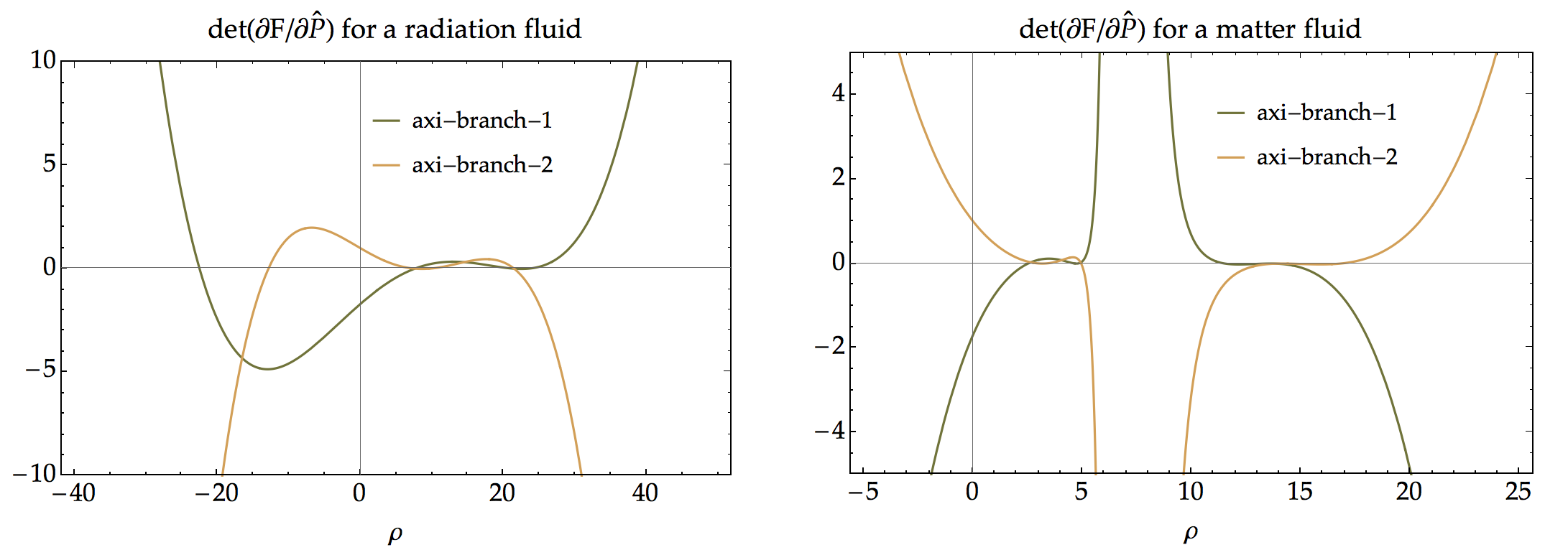} 
    \end{minipage}\hfill
          \caption{Plots of $\det(F_{\hat P})$ for axisymmetric solutions, both with values $\beta=-0.1$ and $\alpha$ is chosen so that $\det(F_{\hat P})_{axi-2}=1$ in vacuum for that value of $\beta$ $(\alpha\approx-0.0213)$. The left plot is for a radiation fluid while the right one is for a matter fluid. } 
    \label{fig:detaniso}
\end{figure}

Let us now turn to the analysis of the axisymmetric solutions, focusing on whether there is any viable mechanism of isotropisation at low densities for any of the branches and sub-branches of solutions to the general quadratic theory. Axisymmetric branches are characterised by $P_{i,j}=P_k$ and $P_j\neq P_i$. We will assume $P_1=P_3\neq P_2$ without loss of generality through this section. As for the isotropic case, there are two branches of anisotropic solutions given by \eqref{aniso1} (axi-branch-1) and \eqref{aniso2} (axi-branch-2). As in the isotropic case, the axi-branch-1 when coupled to a radiation fluid, does not depend on the values of $\alpha$. Concerning the determinant of $F_{\hat P}$ in vacuum, it is independent of the model parameters for axi-branch-1, and takes the same value than in iso-branch-2, suggesting that iso-branch-2 might be an isotropic limit of axi-branch-1. However for axi-branch-2, it does depend on the values of $\alpha$ and $\beta$ as 
\begin{equation}
\lim_{\rho\to0}\det(F_{\hat P})_{axi-2}=\frac{(3 \alpha +5 \beta )^2 (6 \alpha +11 \beta ) \left(6 \alpha ^2+13 \alpha  \beta +9 \beta ^2\right)}{1296 (\alpha +\beta )^3 (2 \alpha +\beta )^2}
\end{equation}
Thus, the parameters could in principle be tuned so that $\det(F_{\hat P})_{axi-2}=1$ in vacuum. Generally, $\det(F_{\hat P})$ has several roots, the number depending on the relations between $\alpha$ and $\beta$ except for the radiation case in axi-branch-1, where it vanishes when $\det(F_{\hat P})\propto(4 \beta  \rho -9) (4 \beta  \rho +3) (4 \beta  \rho +9)^2$. In Fig. \ref{fig:detaniso} we show plots of $\det(F_{\hat P})$ for both branches in the matter and radiation dominated cases.

 As for the properties of the deformation matrix in vacuum, it is complex for axi-branch-1, taking the value $\Omegah_{axi-1}\overset{\rho=0}{=}i\sqrt{3}/2\, {\rm diag}(1,1,-3,1)$,  which is different than that of iso-branch-2, hence implying that one branch cannot be the isotropisation of the other, as neither can be axi-branch-2 due to the dependence on $\alpha$ and $\beta$ of $\Omegah$ in vacuum. This suggests that axisymmetric and isotropic branches are in general non-perturbatively different from isotropic branches for the general quadratic theory even at low densities. However, although it is not possible to find particular combinations of $\alpha$ and $\beta$ such that the deformation matrix becomes the identity in vacuum, we can indeed find particular combinations such that it isotropises in vacuum. Nonetheless, for axi-branch-2, some of its eigenvalues are always negative in vacuum, thus jeopardising the hyperbolic nature of the field equations.  
 \begin{figure}[h]
		\centering
   \begin{minipage}{\textwidth}
        \centering
        \includegraphics[scale=0.45]{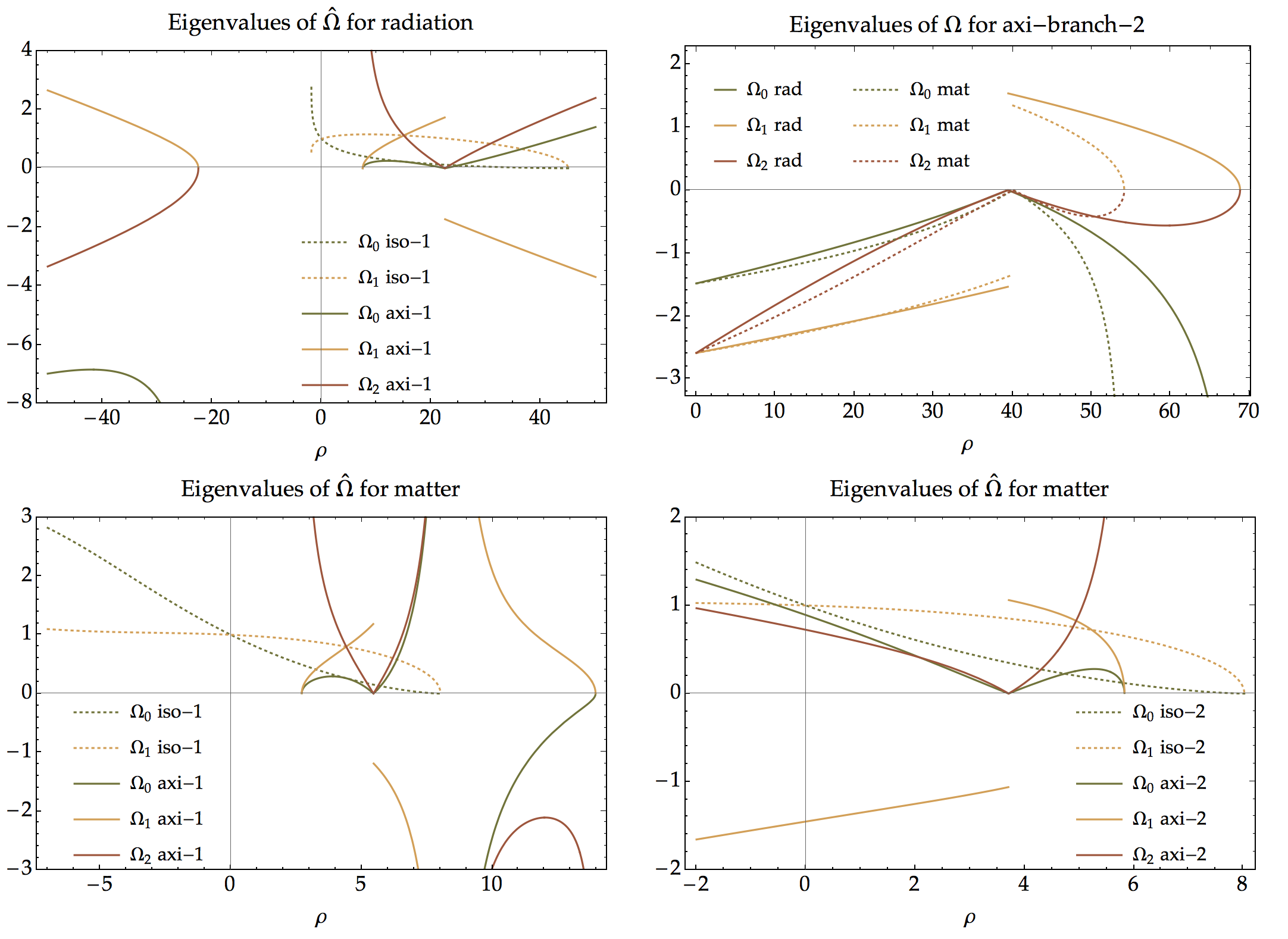}
   \end{minipage}
        \caption{Plots of the eigenvalues of $\Omegah$. The top-left graphic is plotted for the values of $\beta=-0.1$ and the value of $\alpha$ such that $\det(F_{\hat P})_{axi-2}=1$ in vacuum ($\alpha\approx-0.0213$), the top-right is plotted for $\alpha=-0.01$ and the value of $\beta$ such that $\Omegah$ isotropizes in vacuum ($\beta\approx 0.1297$), and both on the bottom are plotted for $\alpha=-0.01$ and $\beta=-0.1$ respectively. The axi-1-branch always isotropizes to iso-branch-1 at some non-zero density for both fluids and in a non-smooth way for the spatial eigenvalues, but the axi-2-branch isotropizes but not to the iso-branch-1 (neither 2) except for a particular value of the parameters. In this case, the spatial eigenvalue does not isotropize at the same value of $\rho$ as the temporal one.}\label{fig:Omegaaxi} 
\end{figure}
Apart from not having a well-defined vacuum, the deformation matrix for axisymmetric solutions is complicated, as can be seen with the examples plotted in figure \ref{fig:Omegaaxi}. There is always a point for which the axi-branch-1 isotropizes and then become anisotropic again as the density grows. At this isotropization point, the eigenvalues of $\Omegah$ of axi-branch-1 coincide with those of iso-branch-1 both for matter and radiation. Nevertheless, the derivatives of the eigenvalues are never the same for isotropic and axisymmetric solutions at that point. The hope that an anisotropic solution could then isotropize in a smooth (and thus predictable) way is in vain. For axi-branch-2, although it isotropizes, it does not meet the iso-branch-1 (remember the only isotropic branch giving the correct low-density limit).

\section{Anisotropic deformation matrix in physical scenarios}\label{sec:PhysicalScenarios}

Having understood which are the necessary conditions for a given RBG theory to have solutions with anisotropic deformation matrix, we can now analyse the consequences in scenarios with physical interest, such as cosmological evolution or black hole scenarios.

\subsection{Application to Cosmology}
\begin{figure}[h]
		\centering
   \begin{minipage}{\textwidth}
        \centering
        \includegraphics[scale=0.45]{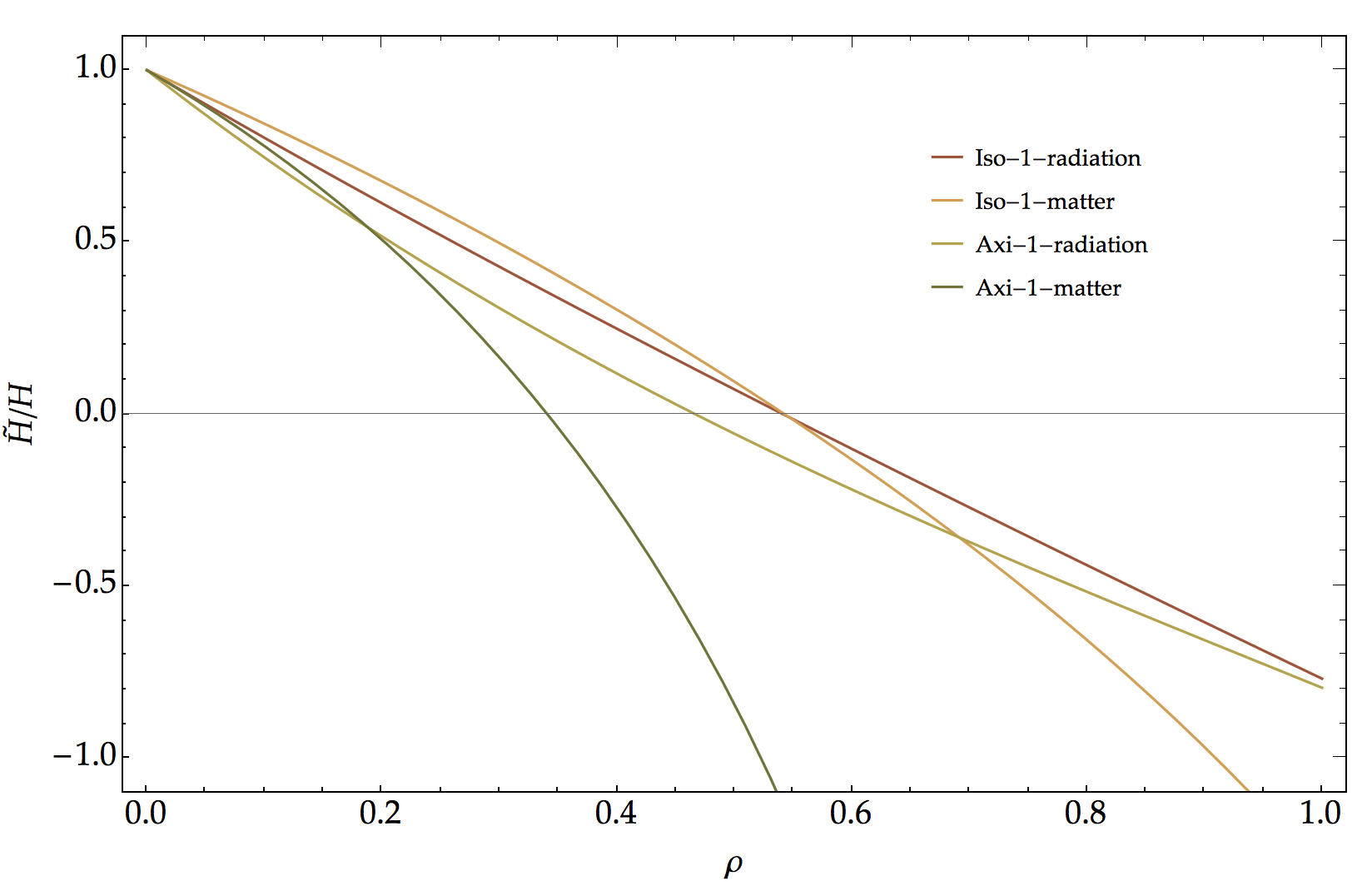}   \end{minipage}
        \caption{Here we plot the ratio between the Hubble factor of the RBG frame ({\it i.e.} that associated to $g_{\mu\nu}$) and the averaged Hubble factor of the Einstein frame ({\it i.e.} that associated to $q_{\mu\nu}$) for the general quadratic theory given by \eqref{Fquadratic}. $\rho$ is normalized by $1/M^2M_{\rm Pl}^2$ and we have chosen $\alpha=-0.2$ and $\beta=-0.1$. We can see how there is a density above which a $g_{\mu\nu}$ expanding phase corresponds to a $q_{\mu\nu}$ contracting phase and viceversa for both isotropic and axisymmetric branches.}\label{fig:quotientHs}
\end{figure}

The results obtained in the previous section apply to general spacetimes filled with a perfect fluid. We will now focus on a cosmological context where the fluid is also homogeneous, i.e., which have a symmetry under spatial translations. Our interest here is to study a scenario where the spacetime metric is isotropic but the $q_{\mu\nu}$ metric is not, so that matter fields do indeed see an isotropic universe but gravitational waves propagate in a non-anisotropic background.\footnote{Recall that minimally coupled matter fields propagate in the background of the spacetime metric in RBG theories, whereas gravitational waves do so according to the background of $q_{\mu\nu}$ \cite{Jimenez:2015caa}.} The spacetime metric will thus have an FLRW form
\be
\d s_g^2=-N^2(t)\d t^2-a^2(t)\d\vec{x}^2
\ee
where we have assumed vanishing curvature of the spatial sections. Since we are exploring solutions where the deformation matrix is not isotropic, the metric $q_{\mu\nu}$ will be of the Bianchi I form
\be
\d s_q^2=-\Nt^2(t)\d t^2-\sum_{i=1}^3a_i^2(t)(\d x^i)^2.
\label{eq:metricqBianchi}
\ee
We can defined the isotropic scale factor $\tilde a=\left[a_1a_2a_3\right]^{1/3}$ and encode the anisotropic expansion in $\gamma_{ij}(t)=e^{2\beta_{i}(t)}\delta_{ij}$, with $\beta_i=\log{(a_i/\tilde a)}$, and no summation over $i$ in the definition of $\gamma_{ij}$ is understood. Notice that the functions $\beta_i$ describing the anisotropic expansion are subject to the constraint
\be\label{betaconstraint}
    \sum_{i=1}^{3}\beta_{i}=0.
\ee
We can now use the relations between $a_i$ and $a$ to define the function $\mA\equiv\tilde a/a=\big(\Omega_1\Omega_2\Omega_3\big)^{1/6}$
that relates the isotropic scale factor of the $q$-metric and the scale factor of $g_{\mu\nu}$. Using this definition, we can write $\beta_i$ in the form
\be
\beta_i=\frac12\log\frac{\Omega_i}{\mA^2},
\ee
and we also have that
\begin{equation}
    \tilde{a}_{i}=\sqrt{\frac{\Omega_{i}}{(\Omega_{1}\Omega_{2}\Omega_{3})^{1/3}}}\hspace{1mm}\tilde{a}=\frac{\sqrt{\Omega_{i}}}{\mathcal{A}}\hspace{1mm}\tilde{a}.
\end{equation}
Furthermore, In Bianchi I, one can define 3 Hubble rates and an averaged one as follows
\begin{equation}
    \tilde{H}_{i}\equiv \frac{\dot{\tilde{a_{i}}}}{a_{i}}=\frac{\diff}{\diff t}\log {\tilde{a_{i}}}\qquad,\qquad \tilde{H}=\frac{\dot{\tilde{a}}(t)}{\tilde{a}(t)}=\frac{\diff}{\diff t}\log {\tilde{a}}(t),
\end{equation}
which by using the continuity equation can be written as
\begin{equation}\label{tildeH}
    \tilde{H}=H\left[1-3(\rho+p)\left(\partial_{\rho}\log \mathcal{A}+c_{s}^{2}\partial_{p}\log \mathcal{A}\right)\right].
\end{equation}
This will allow us to see how when $g_{\mu\nu}$ is in an expanding phase, the metric $q_{\mu\nu}$ can be in a stationary or contracting phase (see fig. \ref{fig:quotientHs}). The above expression is also useful to re-write each of the three Hubble rates in terms of the average variables as
\begin{equation}
    \tilde{H}_{i}=\tilde{H}+\dot{\beta}_{i}.
\end{equation}

It is also convenient to introduce the scalar shear, which measures the deviation from the isotropic case
\begin{equation}
    \sigma^{2}=\sum_{i=1}^{3}\dot{\beta}^{2}.
\end{equation}
One might also wonder what form would have the modified Friedman equation using the metric  \eqref{eq:metricqBianchi}. First we find that the $00$ component of the Einstein tensor of the $q_{\mu\nu}$ metric is
\begin{equation}
    G_{00}(\hat{q})=\tilde{H}_{1}\tilde{H}_{2}+\tilde{H}_{1}\tilde{H}_{3}+\tilde{H}_{2}\tilde{H}_{3}=3\tilde{H}^{2}-\frac{1}{2}\sigma^{2}
\end{equation}
here we have made use of the constraint of Eq. (\ref{betaconstraint}). As we already know, $\tilde{H}$ is given in terms of the energy density and the Hubble rate of the FLRW spacetime metric (see (\ref{tildeH})). Furthermore, the shear can also be expressed in terms of the deformation matrix
\begin{equation}
    \sigma^{2}=\sum_{i=1}^{3}\dot{\beta}^{2}=H^{2}\sum_{i=1}^{3}\left[(\partial_{\rho}\beta_{i}+c_{s}^{2}\partial_{p}\beta_{i})(-3(\rho+p))\right]^{2}
\end{equation}
Therefore,
\begin{equation}\label{G0011}
\begin{split}
      G_{00}(\hat{q})=3H^{2}(t)\Bigg[\left[1-3(\rho+p)\left(\partial_{\rho}\log \mathcal{A}+c_{s}^{2}\partial_{p}\log \mathcal{A}\right)\right]^{2}-\frac{1}{6}\sum_{i=1}^{3}\left[(\partial_{\rho}\beta_{i}+c_{s}^{2}\partial_{p}\beta_{i})(-3(\rho+p))\right]^{2}\Bigg]
\end{split}
\end{equation}
or in terms of of $\hat{P}=\hat{g}^{-1}\hat{R}/\MBIs$. 
\begin{equation}\label{G0022}
    \hat{G}(q)=\hat{R}-\frac{1}{2}\hat{q}\text{Tr}(\hat{q}^{-1}\hat{R})=\MBIs\hat{g}\left[\hat{P}-\frac{1}{2}\hat{\Omega} \text{Tr}(\hat{\Omega}^{-1}\hat{P})\right].
\end{equation}
By means of (\ref{G0011}) and (\ref{G0022}), we obtain the following equation for $H^{2}$
\begin{equation}\label{Hnoniso}
    \frac{3H^{2}}{N^{2}\MBIs}=\frac{\frac12\left(\sum_{i}\frac{\Omega_{0}}{\Omega_{i}}P_{i}-P_{0}\right)}{\left[1-3(\rho+p)\left(\partial_{\rho}\log \mathcal{A}+c_{s}^{2}\partial_{p}\log \mathcal{A}\right)\right]^{2}
    -\frac{1}{6}\sum_{i=1}^{3}\left[(\partial_{\rho}\beta_{i}+c_{s}^{2}\partial_{p}\beta_{i})(-3(\rho+p))\right]^{2}},
\end{equation}
where the right hand side can be written as a function of $\rho$ and $p$ by solving the field equations (\ref{eq:P0}) and (\ref{eq:Pi}). We see that the non-linearities that permit the existence of the anisotropic solutions also complicate the structure of the corresponding Friedman equation. \\

\subsection{Static spherically symmetric geometries}

Another typical scenario would be that of spherically symmetric solutions. We can then study what kind of metric $q_{\mu\nu}$ we can get from an arbitrary static spherically symmetric spacetime metric $g_{\mu\nu}$. As is well known (see {\it e.g.} \cite{Padmanabhan:2010zzb}), a general static and spherically symmetric metric can be written as
\begin{equation}\label{sphericallysymmetric}
ds_g^2=-C(r)dt^2+B^{-1}(r)dr^2+r^2\left(d\theta^2+\sin^2(\theta)d\phi^2\right),
\end{equation}
where $r$ measures the area of the $2-$spheres. Since $\hat\Omega$ can be written in vacuum as an analytic function of $\hat q$ or $\hat g$ and the matter fields, we can assume an arbitrary but diagonalised $\hat\Omega$. Using \eqref{relationqg} and \eqref{sphericallysymmetric} we can then write 
\begin{equation}\label{sphericallysymmetric}
ds_q^2=-\tilde{C}(r)dt^2+\tilde{B}^{-1}(r)dr^2+\tilde{r}^2\left(d\theta^2+\frac{\Omega_3}{\Omega_2}\sin^2(\theta)d\phi^2\right),
\end{equation}
where $\tilde{C}=\Omega_0C$, $\tilde{B}^{-1}=\Omega_1B^{-1}$ and $\tilde{r}^2=\Omega_{2}r^2$ and $r$ can be solved in terms of $\tilde r$ and the matter fields. 

In the presently used coordinates, the metric $q_{\mu\nu}$ is also spherically symmetric in the case that $\Omega_2=\Omega_3$ but otherwise it will be spherically symmetric in some coordinates in which $\phi$ is periodic in $(\Omega_3/\Omega_2)^{-1/2}2\pi$, and thus it will suffer from a conical singularity due to a deficit in angle. Regarding the presence or not of horizons notice that a divergence of the $g_{rr}$ component, which usually shows the presence of event horizons, is also translated as a divergence of the $q_{rr}$ component due to the analiticity of the deformation matrix. Moreover, the divergence takes place for the same value of the $x$ coordinate. Thus, event horizons get mapped in a trivial way.

\section{Anisotropy in the Einstein frame}\label{sec:AnisoEinstein}

After exploring the possibility of having an anisotropic deformation for an isotropic matter source, it is illuminating to look at the problem from the Einstein frame perspective directly. We will briefly summarise the procedure to go to such a frame here that is more extensively discussed in e.g. \cite{BeltranJimenez:2017doy} (see also \cite{Afonso:2017bxr,Jimenez:2020dpn} for an extension to non-minimally coupled matter fields). Since the connection field equations \eqref{connfieldeq} are formally the same as in the first order formalism of GR but here for the metric $q^{\mu\nu}$, the connection can be algebraically solved as the Levi-Civita connection of $q^{\mu\nu}$ (up to an irrelevant projective mode \cite{Bernal:2016lhq}). Taking this into account, the field equations for RBG theories can be recast into \cite{Afonso:2018bpv} 
\be\label{Einstenlikeeqs}
G^\mu{}_\nu(q)=\frac{1}{\mpl^{2}}\tilde{T}^\mu{}_\nu
\ee
where we have defined the Einstein tensor $G^\mu{}_\nu(q)\equiv q^{\mu \rho} R_{\nu \rho}(q)-\frac{1}{2} q^{\rho \sigma} R_{\rho \sigma}(q) \delta^{\mu}{}_{\nu}$ of the metric $q^{\mu\nu}$, and the corresponding energy-momentum tensor
\be\label{EinsteinFrameTmunu}
\tilde{T}^\mu{}_\nu\equiv\frac{1}{\sqrt{\det\Omegah}}\left[g^{\mu\alpha}T_{\alpha\nu}-\left(\mathcal{L}_{G}+\frac{1}{2} T\right) \delta^{\mu}{}_{\nu}\right].
\ee
Since \eqref{relationqg} can be used to algebraically solve $g$ in terms of $q$ and the matter fields, these are nothing but the field equations for GR coupled to the matter sector obtained after integrating out the non-dynamical fields. Indeed, \eqref{relationqg} allows to perform a field redefinition of the metric to find a matter action as a function only of the metric $q_{\mu\nu}$ and the matter fields. This relation is given by 
\begin{equation}
\sqrt{-q} q^{\mu \nu} = 2 \kappa^{2} \sqrt{-g} \frac{\partial F}{\partial R_{\mu \nu}}
\end{equation}
which after writing $F$ in terms of the metric $g_{\mu\nu}$ and the matter fields by means of \eqref{metricfieldeq} gives a non-linear relation between $g_{\mu\nu}$, $q_{\mu\nu}$ and the matter fields analogous to the non-linear equations for $\Omegah$ that we solved for the general quadratic theory in section \ref{sec:GenQuad}. After solving for $\hat g(\hat q, \hat T)$ the original RBG action can be equivalently written as
\be\label{EinsteinFrame}
\mS=\frac12 \mpl^2\int \d^4x\sqrt{-q} q^{\mu\nu} R_{\mu\nu}(q)+\tilde{\mS}_{\rm m}[q_{\mu\nu},\psi],
\ee
where we have collectively denoted the matter fields by $\psi$ and $\tilde{\mS}_{\rm m}$ is the resulting matter action after integrating the connection out and solving for $\hat g(\hat q, \hat T)$. The Einstein equations \eqref{Einstenlikeeqs} then follow from this action upon the following identification:
\be
\frac{-2}{\sqrt{|q|}}q^{\mu\alpha}\frac{\delta\tilde{\mS}}{\delta q^{\alpha\nu}}\equiv\tilde{T}^\mu{}_\nu.
\ee
The action \eqref{EinsteinFrame}, which is equivalent to the RBG action but written in other field variables, is called the Einstein frame of the theory, and we refer to the original form of the theory \eqref{RBGaction} as the RBG frame. It is crucial to realise that no additional dof's have been introduced to go to this frame, thus showing explicitly that the gravitational sector propagates the usual two polarisations of the graviton.

Having arrived at this equivalent formulation of the theory in the usual GR fashion, it is pertinent to ask how to square the obtained anisotropic deformations with the no-hair theorems of GR \cite{Wald:1983ky}. This becomes even more pressing in view of \eqref{EinsteinFrameTmunu} which clearly shows that the source of the Einstein equations for $q_{\mu\nu}$, namely $\tilde T^{\mu}{}_\nu$, is isotropic provided both $T_{\mu\nu}$ and $g_{\mu\nu}$ are. Then, how do we reconcile the general result that the shear decays with the persistent anisotropic solutions obtained in the precedent sections? The resolution to this dichotomy again comes from the non-linearity of the Einstein equations that allows to have anisotropic solutions even if the source is isotropic. The no-hair theorems for cosmological solutions, for instance, states that the anisotropic shear typically decays during the expansion. In our case, we have obtained that it is possible to have an anisotropic deformation, which is equivalent to having a Bianchi I metric for $q_{\mu\nu}$ even if $g_{\mu\nu}$ is of the FLRW type. That the anisotropy can be maintained can be understood from the fact that an expanding solution for the matter fields requires that the metric $g_{\mu\nu}$ describes a growing scale factor, but the evolution for the metric $q_{\mu\nu}$, besides being anisotropic, does not need to correspond to an expanding phase, as can be seen in Fig. \ref{fig:quotientHs}. For instance, if this anisotropic evolution describes a contracting phase, the shear corresponding to $q_{\mu\nu}$ can actually grow substantially while the metric $g_{\mu\nu}$ describes an isotropic expanding phase. On the other hand, even if the evolution also corresponds to an expanding phase, the effective expansion of the metric $q^{\mu\nu}$ can be slower than the one experienced by matter fields so that it can persist after many e-folds of the matter fields expansion.

An interesting example to consider in some detail is that of a cosmological constant or, more generally, matter sectors that are able to support maximally symmetric backgrounds. A quick glance at \eqref{EinsteinFrameTmunu} reveals that a cosmological constant in the RGB frame also gives a cosmological constant in the Einstein frame. If we assume $T_{\mu\nu}=\Lambda g_{\mu\nu}$, then we find that
\be
\tilde{T}^\mu{}_\nu=-\frac{\mathcal{L}_G+\Lambda}{\sqrt{\det\Omegah}}\delta^\mu{}_\nu\equiv\tilde{\Lambda}\delta^\mu{}_\nu.
\ee
By virtue of the Bianchi identities associated to diffeomorphisms, we find that $\tilde{\Lambda}$ must also be a constant so that the solution for $q_{\mu\nu}$ will also correspond to a maximally symmetric metric. It can happen however that a positive $\Lambda$ can lead to a negative or vanishing $\tilde{\Lambda}$. However, drawing any physical conclusion from this is of limited interest since in the absence of propagating matter fields, the only physically relevant object is the metric $q_{\mu\nu}$ that describes the characteristics of the propagation of gravitational waves. In this respect, it should be noticed that what one would call vacuum configuration in the RBG frame is different from the vacuum configuration in the Einstein frame. For instance, if we have a vacuum configuration with $T_{\mu\nu}=0$, in the Einstein frame this configuration would give rise to a cosmological constant. Likewise, if we define the vacuum in the RBG frame as the configuration with trivial matter fields, we can have a cosmological constant, but the value of the cosmological constant in both frames will be different.

The physical effect that could be measured comes when we compare the propagation of gravitational waves and some matter fields. In the minimally coupled case that we are considering, the matter fields follow the geodesics of $g_{\mu\nu}$ while gravitational waves see the metric $q_{\mu\nu}$ (see e.g. \cite{Jimenez:2015caa}). Let us assume that $g_{\mu\nu}=\eta_{\mu\nu}$ and $\hat{\Omega}$ is anisotropic so we have $q_{\mu\nu}={\rm diag}(N,a,b,c)$ and, for simplicity, we will assume that they are constant (i.e. we are considering vacuum configurations). If we now compare the trajectories of photons and gravitons, they respectively follow the null geodesics of the metrics:
\begin{eqnarray}
&&\d s^2_g=-\d t^2+\d \vec{x}^2,\\
&&\d s^2_q=-N\d t^2+a\d x^2+b\d y^2 +c\d z^2.
\end{eqnarray}
If we emit a graviton and a photon at $t=t_0$ from the origin along the $z-$direction, we will have
\be
z_{\rm photon}=t-t_0,\quad z_{\rm graviton}=\frac{N}{c}(t-t_0)
\ee
so their trajectories differ as $\Delta z= \left(1-\frac{N}{c}\right)(t-t_0)$. This would of course be tightly constrained by the observations of the neutron star merger \cite{TheLIGOScientific:2017qsa}. An important point to realise is that the effect of the anisotropic $\hat{\Omega}$ cannot be absorbed into a coordinate redefinition, since that would affect the propagation of the matter fields and the relative separation would remain. In the standard case, the fact that all fields follow the same metric is what allows to absorb the anisotropic solutions of vacuum Einstein equations  that we have considered into a redefinition of the coordinates so that it does not have any physical effect. Furthermore, notice that this effect does not depend on the deformation matrix being anisotropic, but it will arise whenever $\hat{\Omega}\neq\Id$. The fact of having an anisotropic deformation matrix will further introduce polarisation and direction dependent effect. \\

Let us end our discussion on the Einstein frame by explaining another subtle point that usually arises when going to this frame. This subtlety is related to need to solve the non-linear equation for the deformation matrix that has been the core of this work. The Einstein frame formulation of the RBG theories can be achieved directly working at the level of the equations, in which case one ends up with Eq. \eqref{Einstenlikeeqs}. In those equations, the right hand side depends on the metric $g_{\mu\nu}$ so, in order to properly have the differential equations determining $q_{\mu\nu}$, one needs to solve the equation for the deformation matrix $\Omega^\mu{}_\nu$. It is then usually assumed that the solution can be written as a covariant expression of the stress-energy tensor. By virtue of the Cayley-Hamilton theorem, one is then entitled to make the Ansatz
\be
\hat{\Omega}=\sum_{n=0}^3 c_n\hat{T}^n
\label{eq:OmegaTn}
\ee
with $c_n$ some scalar functions of the invariants of $T^\mu{}_\nu$ and $\hat{T}^n$ denotes the $n$-th power. However, though this is a very reasonable and natural guess, it does not cover the full space of solutions. This should be clear from our results above and, owed to the non-linear nature of the matrix equation satisfied by $\hat{\Omega}$, more general solutions are possible where the explicit covariant relation exhibited in \eqref{eq:OmegaTn} is spontaneously broken. For example, in vacuum, one can have solutions where $\hat{\Omega}$ is not proportional to the identity so that Lorentz invariance is spontaneously broken. The same can happen for non-vacuum situations. In the construction of the Einstein frame at the level of the action directly, the same situation occurs when one has to integrate out the metric $g_{\mu\nu}$. Again, this is done by solving its algebraic equation, which is non-linear and allows for branches of solutions that do not explicitly preserve covariance. After plugging these solutions in the action, the matter sector will then contain the effects of those non-trivial branches.

\section{Discussion}\label{sec:Discusion}
In this work we have unleashed the common assumption within RBGs of an isotropic deformation and explored the possibility of having anisotropic deformations in the presence of isotropic matter. We have studied the general conditions for a given theory to be able to accommodate anisotropic deformations. From the resulting general condition, we have unveiled the noteworthy property of EiBI theories (and some of its extensions) that there are no anisotropic deformations in presence of isotropic matter. However, for more general theories, this is not the case and anisotropic deformations in presence of isotropic matter are (in general possible). We have studied in more detail some specific theories and, in particular, we have exhaustively analised the general quadratic theory. For these, we have obtained that, even though some branches of solutions correspond to anisotropic deformations, they are generically pathological at low densities where the branches do not exist. Obviously, these branches are disconnected from the solution that continuously connects with GR at low densities. Despite the specificity of this result for quadratic theories, it makes apparent that branches of any theory that are perturbatively close to GR at low densities, do not admit smooth anisotropic deformations. Thus, the anisotropic branches of more general theories with a smooth behaviour at low densities must be non-perturbative, i.e., they must strongly rely on its non-linear nature. As applications of our general formalism, we have considered the case of cosmological scenarios and spherically symmetric spacetimes. Finally, we have discussed how the obtained results can be understood from the perspective of the Einstein frame that these theories admit in terms of the evolution of the shear.

In conclusion, the usual isotropic Ansatz, besides being a natural choice, it may be necessary to avoid pathologies. We should notice however that the suitability of the isotropic deformation was not guaranteed a priori. As an example we can mention the cosmological isotropic bouncing solutions that can be unstable due to the growth of the shear in the contracting phase and something along these lines (barring the obvious differences) may have happened for the solutions with isotropic deformation in RBGs. Our analysis then provides a strong support for the physical motivation of the isotropic Ansatz. \\

\acknowledgments
It is a pleasure to thank Gonzalo Olmo and Diego Rubiera-Garc\'ia for useful discussions and comments.  JBJ acknowledges support  from  the Atracci\'on del  Talento  Cient\'ifico en  Salamanca  programme and the  project  PGC2018-096038-B-I00  by Spanish Ministerio de Ciencia, Innovaci\'on y Universidades. AD is supported by  the fellowship FPU15/05406  (MINECO). This work is supported by the the Spanish Projects No. FIS2017-84440-C2-1-P (MINECO/FEDER, EU), the Project No. H2020-MSCA-RISE-2017 Grant No. FunFiCO-777740, Project No. SEJI/2017/042 (Generalitat Valenciana), the Consolider Program CPANPHY-1205388, and the Severo Ochoa Grant No. SEV-2014-0398 (Spain). This article is based upon work from COST Action CA15117, supported by COST (European Cooperation in Science and Technology). AD also wants to thank hospitality to the \textit{Departamento de F\'isica Te\'orica de la Universidad de Salamanca}

\bibliographystyle{JHEPmodplain}

\bibliography{Bibliography}

\end{document}